\documentclass[aps,twocolumn,prb,superscriptaddress]{revtex4-2}

\usepackage{dcolumn}
\usepackage{bm}
\usepackage{bbold}

\usepackage[utf8]{inputenc}
\usepackage[english]{babel}
\usepackage{mathtools}
\usepackage{amsfonts}
\usepackage{mathrsfs}
\usepackage{bbm}
\usepackage{slashed}
\usepackage{tensor}

\usepackage{tikz}
\usetikzlibrary{positioning}
\usepackage{graphicx}
\usepackage{color}
\usepackage{array}
\usepackage[abs]{overpic}

\usepackage{placeins}

\usepackage{makecell}

\usepackage{xspace}
\usepackage{siunitx}
\usepackage{xfrac}
\usepackage{hyperref}
\usepackage[nameinlink]{cleveref}
\usepackage{appendix}
\usepackage{xifthen}
\usepackage{xcolor}
\hypersetup{
	colorlinks,
	linkcolor={red!75!black},
	citecolor={blue!75!black},
	urlcolor={blue!75!black}
}

\newcommand{\sbra}[1]{\langle #1 |}
\newcommand{\sket}[1]{| #1 \rangle}

\usepackage{tikz}

\newcommand{\urmove}[0]{%
\begin{tikzpicture}[baseline=0.1ex]%
\draw[thick, red] (0,1.5ex) -- (1.5ex,1.5ex);%
\draw[thick, red] (1.5ex,1.5ex) -- (1.5ex,0);%
\end{tikzpicture}%
}

\newcommand{\ulmove}[0]{%
\begin{tikzpicture}[baseline=0.1ex]%
\draw[thick, red] (0,1.5ex) -- (1.5ex,1.5ex);%
\draw[thick, red] (0,1.5ex) -- (0,0);%
\end{tikzpicture}%
}

\newcommand{\dlmove}[0]{%
\begin{tikzpicture}[baseline=0.1ex]%
\draw[thick, red] (0,1.5ex) -- (0,0);%
\draw[thick, red] (0,0) -- (1.5ex,0);%
\end{tikzpicture}%
}

\newcommand{\drmove}[0]{%
\begin{tikzpicture}[baseline=0.1ex]%
\draw[thick, red] (0,0) -- (1.5ex,0);%
\draw[thick, red] (1.5ex,1.5ex) -- (1.5ex,0ex);%
\end{tikzpicture}%
}

\newcommand{\vvmove}[0]{%
\begin{tikzpicture}[baseline=0.1ex]%
\draw[thick, red] (0,0) -- (0,1.5ex);%
\draw[thick, red] (1.5ex,0) -- (1.5ex,1.5ex);%
\end{tikzpicture}%
}

\newcommand{\hhmove}[0]{%
\begin{tikzpicture}[baseline=0.1ex]%
\draw[thick, red] (0,0) -- (1.5ex,0);%
\draw[thick, red] (1.5ex,1.5ex) -- (0,1.5ex);%
\end{tikzpicture}%
}

\usepackage{booktabs}
\usepackage{multirow}

\newcolumntype{C}{>{$}c<{$}}
\AtBeginDocument{
	\heavyrulewidth=.08em
	\lightrulewidth=.05em
	\cmidrulewidth=.03em
	\belowrulesep=.65ex
	\belowbottomsep=0pt
	\aboverulesep=.4ex
	\abovetopsep=0pt
	\cmidrulesep=\doublerulesep
	\cmidrulekern=.5em
	\defaultaddspace=.5em
}

\graphicspath{{./figures/}}

\newcommand{\gettitle}{Time evolution of the quantum Ising model in two dimensions \\ using Tree Tensor Networks}

\newcommand{\getJuelichAffiliation}{\affiliation{Institute of Quantum Control (PGI-8), Forschungszentrum Jülich, D-52425 Jülich, Germany}}
\newcommand{\getRegensburgAffiliation}{\affiliation{Faculty of Informatics and Data Science, University of Regensburg, D-93040 Regensburg, Germany}}
\newcommand{\getKoelnAffiliation}{\affiliation{Institute for Theoretical Physics, University of Cologne, D-50937 Köln, Germany}}
\newcommand{\getAugsburgAffiliation}{\affiliation{Theoretical Physics III, Center for Electronic Correlations and Magnetism, Institute of Physics, University of Augsburg, D-86135 Augsburg, Germany}}
\newcommand{\getCAAPSAffiliation}{\affiliation{Centre for Advanced Analytics and Predictive Sciences (CAAPS), University of Augsburg, Universitätsstr. 12a, 86159 Augsburg, Germany}}

\hypersetup{
	pdftitle={\gettitle},
	pdfauthor={},
	bookmarksopen=true,
	bookmarksopenlevel=2,
	bookmarksnumbered=true
}

\begin{document}

\title{\gettitle}

\author{Wladislaw Krinitsin}\getJuelichAffiliation\getRegensburgAffiliation
\author{Niklas Tausendpfund}\getJuelichAffiliation\getKoelnAffiliation
\author{Markus Heyl}\getAugsburgAffiliation\getCAAPSAffiliation
\author{Matteo Rizzi}\getJuelichAffiliation\getKoelnAffiliation
\author{Markus Schmitt}\getJuelichAffiliation\getRegensburgAffiliation

\begin{abstract}
The numerical simulation of two-dimensional quantum many-body systems away from equilibrium constitutes a major challenge for all known computational methods.
We investigate the utility of Tree Tensor Network (TTN) states to solve the dynamics of the quantum Ising model in two dimensions. 
Within the perturbative regime of small transverse fields, TTNs faithfully reproduce analytically known, but non-trivial and physically interesting dynamics of magnetic domains, for lattices up to $16 \times 16$ sites. Limitations of the method related to the rapid growth of entanglement entropy are explored within more general, paradigmatic quench settings. 
We provide and discuss comprehensive benchmarks regarding the benefit of GPU acceleration and the impact of using local operator sums on the performance.
\end{abstract}
\maketitle

\section{Introduction}

Tensor network algorithms serve as potent and versatile numerical methods in many-body quantum physics, enabling the computation of static and dynamic properties of generic lattice systems~\cite{Orus_2014,Paeckel_2019,silvi_2019,Cirac_2021}. 
These algorithms achieve efficiency by compressing quantum states into a network of low-rank tensors with bonds of a fixed dimension, turning the exponential growth of the Hilbert space with the system size into a polynomial one. 
The limitations of such a compressed representation are well understood -- it is constrained by the states' entanglement complexity.
For one-dimensional, local and gapped systems the entanglement entropy follows the boundary law, enabling ground state search using the highly successful \textbf{D}ensity \textbf{M}atrix \textbf{R}enormalization \textbf{G}roup (DMRG) algorithm~\cite{White_1992}.
Immediate generalizations to higher dimensions suffer from higher algorithmic complexity and typically less favorable entanglement scaling.
Nonetheless, recent advances in the area of believe propagation~\cite{Tindall_2023,Tindall_2024} and \textbf{P}rojected \textbf{E}ntangled \textbf{P}air \textbf{S}tates (PEPS)~\cite{Verstraete_2004,Jordan_2008,Lubasch_2014,Lubasch_2014_2,Corboz_2016,Vanderstraeten_2016,Zaletel_2020,Puente_2025} provide a promising avenue for efficiently representing ground states in higher dimensional quantum systems~\cite{Murg_2007,Murg_2009,Niesen_2017,Liao_2017,Chen_2018,Hasik_2022,Hasik_2024,Weerda_2024}.

With \textbf{N}eural \textbf{Q}uantum \textbf{S}tates (NQS)~\cite{Carleo_2017}, an alternative family of variational states emerges, whose neural network representation of the wave function offers higher flexibility concerning the lattice geometry of the physical system. While their capability to capture volume law entanglement \cite{deng_quantum_2017,levine_quantum_2019,Denis_2025} and first results for the dynamics of two dimensional quantum systems~\cite{Schmitt_2020,Gutierrez_2022,schmitt_quantum_2022,Sinibaldi_2023,mendes-santos_highly_2023,mendes-santos_wave-function_2024,Vandewalle_2024,Sinibaldi_2024} outline a promising avenue, their limitations and shortcomings are not yet fully understood.
Therefore, tackling dynamics in two-dimensional systems remains a major challenge for all computational methods.

On the tensor network side, a less explored architecture are so called \textbf{T}ree \textbf{T}ensor \textbf{N}etworks (TTNs) -- their hierarchical structure can be exploited to respect the local nature of interactions present in typical Hamiltonians, while still remaining loop-free and generalizable to lattices of any dimension.
So far they have been successfully applied to the ground state search~\cite{Tagliacozzo_2009,Murg_2010,Ferris_2013,Schuhmacher_2025} and recently also used to tackle questions related to the dynamics of two-dimensional quantum systems~\cite{Schroder_2019,kloss_2020,Pavesic_2024,Jaschke_2024}.

In the following we expand upon these results, focusing on the example of the quantum Ising model, which serves as a paradigmatic model in the study of quantum phase transitions and other collective phenomena in and out of equilibrium.
In recent years, it has become highly relevant as the effective model describing the main interactions in Rydberg atom arrays~\cite{Greiner_2002,Kinoshita_2006,Hild_2014,Georgescu_2014,Martinez_2016,Jae-yoon_2016,Gross_2017,Jurcevic_2017,Bernien_2017,Zhang_2017,Gaerttner_2017,Choi_2017,Levine_2018,Barredo_2018}.
With local addressibility becoming an increasingly common capability of such quantum simulators, specific inhomogeneous product states can be prepared in order to observe their relaxation dynamics~\cite{Manovitz2024}.
This opens the possibility to investigate interesting theoretical predictions.
In particular, we will address various domain wall initial conditions, which exhibit different inhibiting mechanisms, that result in unusually slow thermalization.
On the one hand, approximate dynamical constraints lead to Hilbert space fragmentation in the perturbative limit of small transverse fields, leading to effecive Wannier-Stark localization of interfaces and extremely long-lived pre-thermal states~\cite{Balducci_2022,balducci_interface_2022}.
Away from the perturbative regime, the stability of domain wall initial conditions persists as long as the system remains within a smooth interface regime below a roughening cross-over~\cite{Krinitsin2024}.
The associated slow growth of entanglement renders tensor network techniques a suited tool to investigate the characteristics of such dynamics.

In this work we investigate the capabilities of TTNs in regimes of constrained dynamics and paradigmatic quench settings. We provide a comprehensive study of the numerical method, including detailed performance benchmarks.
We start by recapitulating the general concepts behind TTNs in \Cref{sec:ttns} and the \textbf{T}ime-\textbf{D}ependent \textbf{V}ariational \textbf{P}rinciple (TDVP) used to perform the time evolution in \Cref{sec:tdvp}. 
Benchmarks and discussions of strategies enhancing the performance of the algorithm used in this work, like the usage of local sums instead of \textbf{M}atrix \textbf{P}roduct \textbf{O}perators and the performance gain on \emph{GPU}s, are provided in \Cref{sec:performance_strategies}.
We introduce the transverse field Ising model and analytical predictions related to its perturbative limit in \Cref{sec:tfim}, after which we turn to a numerical  investigation of said predictions for various classes of initial states in \Cref{sec:results}. 
Away from the perturbative limit, we provide an analysis for quenches of different strengths, pointing out limitations of the method.
A summary of the main results and outlook in \Cref{sec:discussion} conclude this work.

\section{Numerical and theoretical Details}

\subsection{Tree Tensor Networks} \label{sec:ttns}
\begin{figure}[hbt!]
    \centering
    \includegraphics[width=0.9\linewidth]{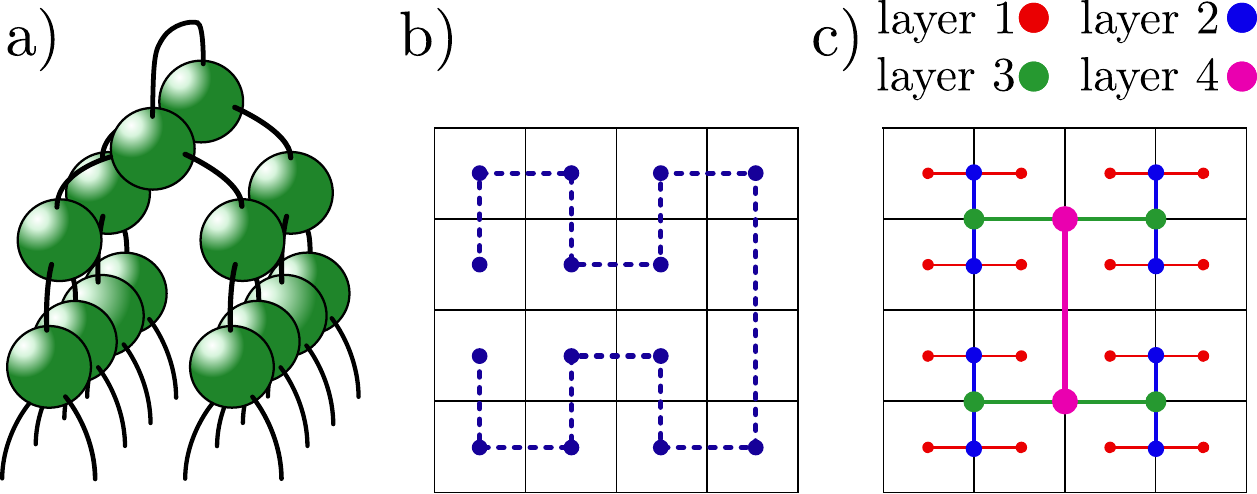}
    \caption{a) Binary TTN architecture on top of a two dimensional, physical lattice. b) Covering of the two dimensional lattice by the most optimal Hilbert-curve. c) Contraction directions for each layer within the TTN, alternating between x and y from layer to layer. Such a construction traces the Hilbert-curve, introduced in b), increasing the covered patches of the lattice when going up the layers.} \label{fig:ttn}
\end{figure}

Tree tensor networks provide a highly versatile, loop-free network architecture, which makes them easily applicable to arbitrary lattice geometry while keeping the computational complexity tractable. By alternating the direction of inter-layer connections, i.e. for a two dimensional lattice alternating between connections oriented along the x- and y-direction, see Fig.~\ref{fig:ttn}a), the tree structure is able to provide a coverage of higher-dimensional lattices, in which the average path distance through the network between two physical sites remains constant in the thermodynamic limit~\cite{Felser_2021}.
Thus, fewer non-local interactions within the network are introduced, as opposed to the inherently one-dimensionally structured \textbf{M}atrix \textbf{P}roduct \textbf{S}tates (MPS).
In the following, we provide a summary of the technicalities related to tree tensor networks, following Refs.~\cite{Shi_2006,Tagliacozzo_2009,Murg_2010,kloss_2020,Felser_2021}.

In this work, we consider binary TTNs, i.e. trees where each node in layer $l$ is connected to two \emph{child} nodes in layer $l-1$, see Fig.~\ref{fig:ttn}a for a graphical representation of the network and Fig.~\ref{fig:ttn}b for a visualization of the node-connectivities. 
The node shared by these two child nodes is being denoted as their parent node. The nomenclature in this work is such that physical legs are attached to tensors at the bottom, i.e. the $l=1$ layer. 

The computational cost of TTN algorithms is dominated by matrix decompositions (singular value decomposition (SVD) or QR decomposition) and tensor contractions.
To make calculations tractable, a finite, fixed bond dimension $\chi$ determines the dimensionality of links connecting tensors sitting on nodes within the network. 
This results in a computational cost of $\mathcal{O}(N \log N \chi^4)$ for both the matrix decompositions and tensor contractions, and memory cost of $\mathcal{O}(N \log N \chi^3)$ for binary TTNs with $N$ physical legs.
The limiting factor regarding the representability power of a tensor network is given by the entanglement spectrum of the underlying physical state; 
the bond dimension determines how many singular values $\sigma_a$ are being taken into account, which in turn determines the maximal representable bi-partite entanglement entropy via the relation $S_\mathrm{ent} = -\sum \sigma_a^2 \log \sigma_a^2$. 
Therefore, TTNs constitute a flexible ansatz to efficiently represent two-dimensional quantum states with limited bi-partite entanglement.

Due to its structure, the orientation of a TTN with respect to the underlying physical lattice may play an important role. 
While bonds up to level $l$ s.t. $2^l<\chi$ are always saturated, the top bond, connecting the two halves of the lattice, suffers most from the truncation to a finite bond dimension and can thus introduce a large bias to the calculation.
Tracking the convergence of the entanglement entropy with respect to the bond dimension at different levels of the tree can be used to understand the entanglement structure and potential bottlenecks in the representation of the quantum state. 
Furthermore, rotating the tree structure with respect to the underlying lattice provides an additional check regarding potential biases in the representation, which can be especially important in the case of non-isotropic systems.

\begin{figure}[t!]
    \centering
    \includegraphics[width=0.9\linewidth]{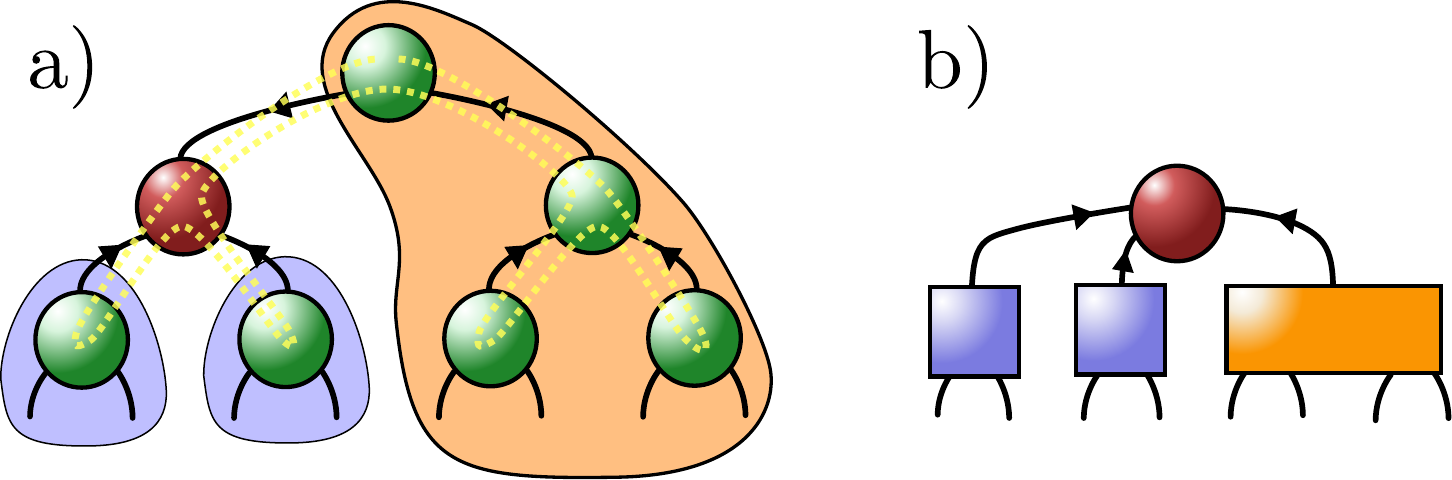}
    \caption{a) The red tensor represents the orthogonality center, the tensors contained withing the blue/orange shaded area can be contracted to represent its children/parent environments, respectively. The yellow dotted lines traces the path in which the tensors are updated when performing the TDVP algorithm, introduced in Sec.~\ref{sec:tdvp}. b) A compact notation for tensor and its environments.} \label{fig:envs}
\end{figure}

Analogously to MPS, tree tensor networks have a gauge freedom which does not change the physical state. 
This gauge freedom can be utilized in order to define an orthogonality center, towards which all other nodes in the network are isometrized, see Fig.~\ref{fig:envs}. 
This is achieved iteratively, by performing sequential QR decompositions or SVDs on all nodes, starting from the ones farthest away within the network. 
One of the unitaries resulting from the SVD/QR replaces the respective tensor, while the remainder is being multiplied onto the next one, in the direction of the desired isometry center, graphically represented by an arrow.
This process is repeated until the orthogonality center is reached; at that point all tensors except the one sitting at the orthogonality center are isometries.

Having performed such an isometrization w.r.t. the node $A_{i,l}$ at position $i$ within layer $l$, one defines parent and children environments as shown in Fig.~\ref{fig:envs} by contracting all isometries within branches starting from the child and parent nodes of node $A_{i,l}$, respectively. 
By tracking the environments of each node, this isometrization can be used to formulate very efficient, iterative algorithms such as the TDVP.

\subsection{Time Dependent Variational Principle} \label{sec:tdvp}

The TDVP is a method to perform time evolution on some parameterized ansatz, since it restricts the dynamics from the full Hilbert space to some generalized phase space, 
where the motion is described by classical Poisson brackets. 
It arises from minimizing the Lagrangian~\cite{Haegeman_2011,Haegeman_2016,Paeckel_2019}
\begin{equation}
    \mathcal{L} = \langle \Psi_\theta | i\partial_t - \hat{H} | \Psi_\theta \rangle
\end{equation}
with respect to the variational parameters $\theta$, where $\hat{H}$ denotes the Hamiltonian of interest. The resulting equation of motion is given by
\begin{equation}
    i\partial_t | \Psi_\theta \rangle = \mathcal{P}_T(\Psi_\theta) \hat{H} | \Psi_\theta \rangle, \label{eq:TDVP_eq1}
\end{equation}
with $\mathcal{P}_T(\Psi_\theta)$ the projector onto the tangent space of the variational manifold at position $\Psi_\theta$. By performing an additive splitting of the projector and a suitable gauge choice~\cite{Ceruti_2020,Bauernfeind_2020,kloss_2020}, solving \eqref{eq:TDVP_eq1} reduces to the following equations for each node tensor $A^{[l,i]}$ and link tensor $L^{[l,i]}$
\begin{align}
    i \dot{A}^{[l,i]} &= + H^{[l,i]}_\mathrm{eff} A^{[l,i]}, \nonumber\\
    i \dot{L}^{[l,i]} &= - \tilde{H}^{[l,i]}_\mathrm{eff} L^{[l,i]}, \label{eq:TDVP_eq2}
\end{align}
where the effective Hamiltonians is defined as the contraction of the full Hamiltonian and environments of the local tensor $A^{[l,i]}$. Note the minus sign in the equation of the link tensor, which can be understood as evolving it backward in time.
Given a small enough timestep $\Delta t$, the solution to Eq.~\eqref{eq:TDVP_eq2} is given by
\begin{align}
    A^{[l,i]} &= \exp^{-i H^{[l,i]}_\mathrm{eff} \Delta t} A^{[l,i]}, \nonumber\\
    L^{[l,i]} &= \exp^{+i \tilde{H}^{[l,i]}_\mathrm{eff}\Delta t} L^{[l,i]}. \label{eq:TDVP_eq3}
\end{align}
We employ a Krylov-based solver to perform the time propagation, which is limited by the cost of repeatedly applying the effective Hamiltonian to the tensor at question. The maximum number of Krylov steps is set to $30$ throughout the simulation.

The order in which the tensors are updated is related to the resulting time-step error; the path used in this work is split into a counter- and clockwise walk, tracing the yellow dotted line shown in, Fig.~\ref{fig:ttn}a) -- starting from the top node, tensors are evolved in time using a half-time step when moving up (down) a layer while tracing the path (counter-)clockwise, otherwise only the orthogonality center is moved~\cite{kloss_2020}. The resulting error with respect to the chosen time step $\delta t$ scales as $\mathcal{O}(\delta t ^3)$.

In the next section we discuss different approaches related to increasing the performance of the algorithm, first by analyzing the impact of using GPUs over CPUs, and afterwards how to efficiently construct and evaluate these effective Hamiltonians using the local sum formalism.
\subsection{Performance Strategies} \label{sec:performance_strategies}
\subsubsection{GPU vs CPU}
\begin{figure}[t!]
    \centering
    \includegraphics[width=0.9\linewidth]{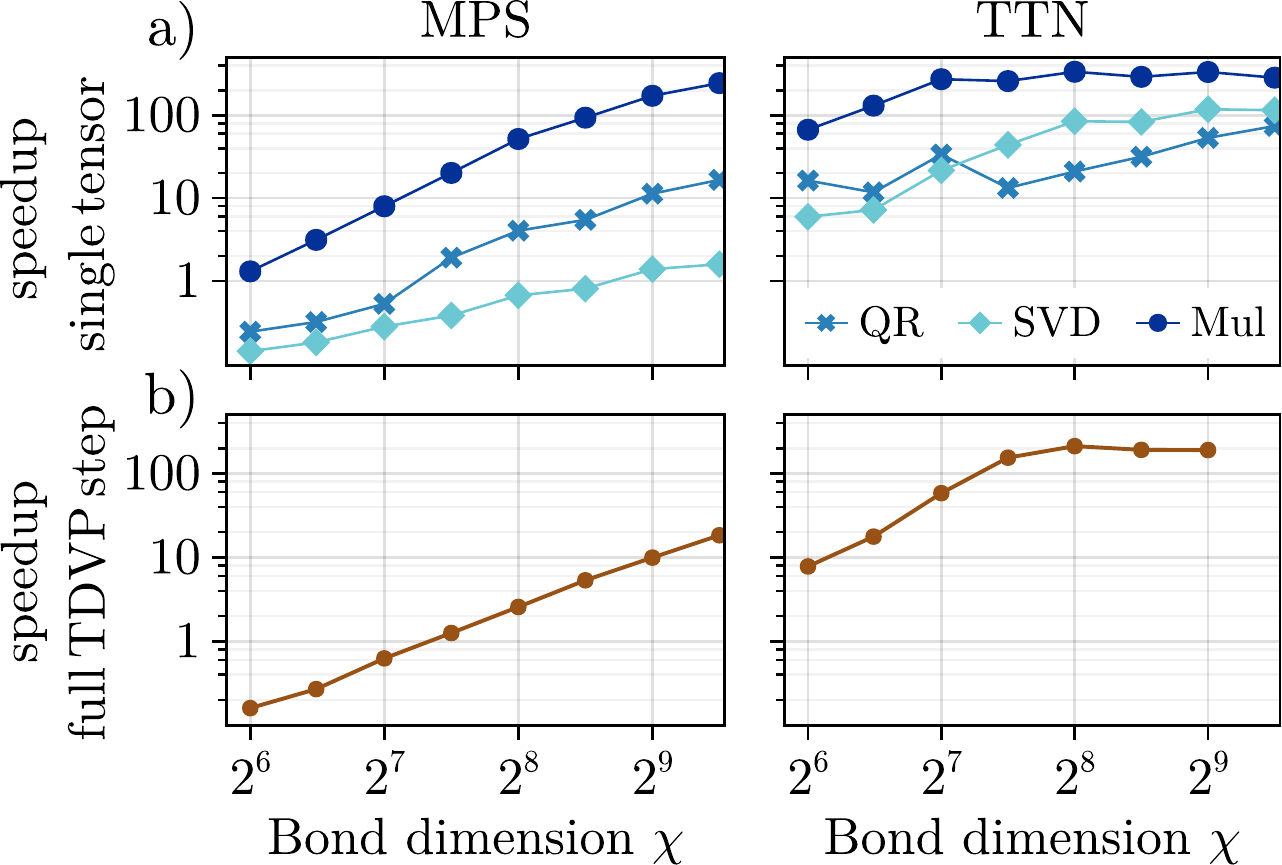}
    \caption{a) GPU vs CPU (single-threaded) speedup of the QR/SVD-decompositions and tensor multiplication applied onto single tensors of size $\chi \times \chi$ (as they appear in an MPS) and $\chi^2 \times \chi$ (as in a binary TTN). b) GPU vs CPU (single-threaded) speedup of a full TDVP step applied to an MPS (left) and TTN (right) at max bond dimension $\chi$.} \label{fig:timings}
\end{figure}

The computational cost of tensor contractions as well as of tensor decompositions is of order $\mathcal O(\chi^{4})$ as a function of the bond dimension $\chi$. Tensor contractions are highly parallelizable, in contrast to tensor decompositions in general.
However, due to the tree structure the decomposition is being performed on so-called tall-skinny matrices of size $\chi^2 \times \chi$ for which batched versions of tensor decomposition algorithms exist~\cite{Anderson_2011,Kerr_2009}, providing a substantial speedup when these operations are being performed on the GPU. 
To highlight the relevance of the tensor dimensions, we compare the acceleration of MPS and TTN operations.
Fig.~\ref{fig:timings}a) provides said comparison using an \emph{NVIDIA A100} GPU and an \emph{Intel Xeon Gold} 6240 CPU ($2.60\, \mathrm{GHz}$) for the QR-/SVD-decomposition as well as the tensor multiplication, on tensors of size $\chi \times \chi$ (left), as they are typically found in MPS, and tensors of size $\chi^2 \times \chi$ (right), as they occur in the binary TTN. 

For the MPS-type tensors, the speedup on a single-threaded process for the tensor multiplication, as expected, is on the order of $\mathcal O(10^2)$. The tensor decompositions seem to behave differently with regards to their parallelization: while we don't observe any speedup for the SVD, performing the QR-decomposition on a GPU provides a speedup of around $10\times$ for the largest bond dimensions considered.
As a consequence, the speedup we observe for performing a single-site TDVP step, i.e. evolving all tensors of the tree by one timestep, together with the SVD as the tensor decomposition algorithm on a MPS with $64$ sites goes up to $20\times$ as the bond dimension increases, see Fig.\ref{fig:timings}b). 
This speedup can be explained by the fact that the tensor multiplication is being performed much more often than the SVD due to the repeated application of the effective Hamiltonian to tensors as part of the Krylov solver. 
By replacing the SVD through a QR-decomposition in situations where a truncation of singular values is not needed, e.g. single-site schemes, the speedup can be increased, potentially by an order of magnitude.

For TTN-type tensors on the other hand, the speedup for all operations is of the order of $\mathcal O(10^2)$. As a consequence, the overall simulation, i.e. performing a TDVP step, profits substantially from GPU acceleration - we observe a time speedup of around $200\times$ over the CPU performance on a TTN (Fig.\ref{fig:timings}b). 
We want to point out that enabling multithreading on the CPU with 24 cores will reduce the observed speedup, e.g. from $200\times$ to around $70\times$ for performing a TDVP step with TTNs. 
However, since that introduces various other contributions like caching etc, impacting the scaling, we opted for a comparison using single-threaded processes to better disentangle the different effects.

Nonetheless, even the speedup of $70 \times$ is substantial, as the most expensive simulation at a bond dimension $\chi=362$, with a time step of $dt=0.1$ up to $t_\mathrm{max}=100$ will take around two days on a single \emph{A100} GPU -- on a CPU (\emph{Xeon Gold}), with multithreading enabled, the same calculation takes around 145 days.
\subsubsection{Matrix Product Operator vs Local Sums}

A typical choice when solving one-dimensional systems using tensor networks is to represent the Hamiltonian as a \textbf{M}atrix \textbf{P}roduct \textbf{O}perator (MPO)~\cite{VanDamme_2024} of bond dimension $K$. 
Because of the existence of a natural order in one dimension, every Hamiltonian with a finite interaction range can be represented by a MPO with a fixed bond dimension, independent of the system size, allowing an efficient evaluation of~eq.~\eqref{eq:TDVP_eq3}.

In two dimensions this is not the case anymore, as one has to define a space filling curve mapping the two dimensional space onto a chain. 
Even by using the most optimal choice, the Hilbert-curve~\cite{Cataldi_2021}, see Fig.~\ref{fig:ttn}b, for a square lattice of dimensions $L \times L$ the interactions have a distance scaling as $\mathcal{O}(L^2)$ in the linear index of the chain.
This results in a system size dependent bond dimension, which renders simulations of large systems unfeasible using the MPO ansatz.

\begin{figure}[t!]
    \centering
    \includegraphics[width=0.9\linewidth]{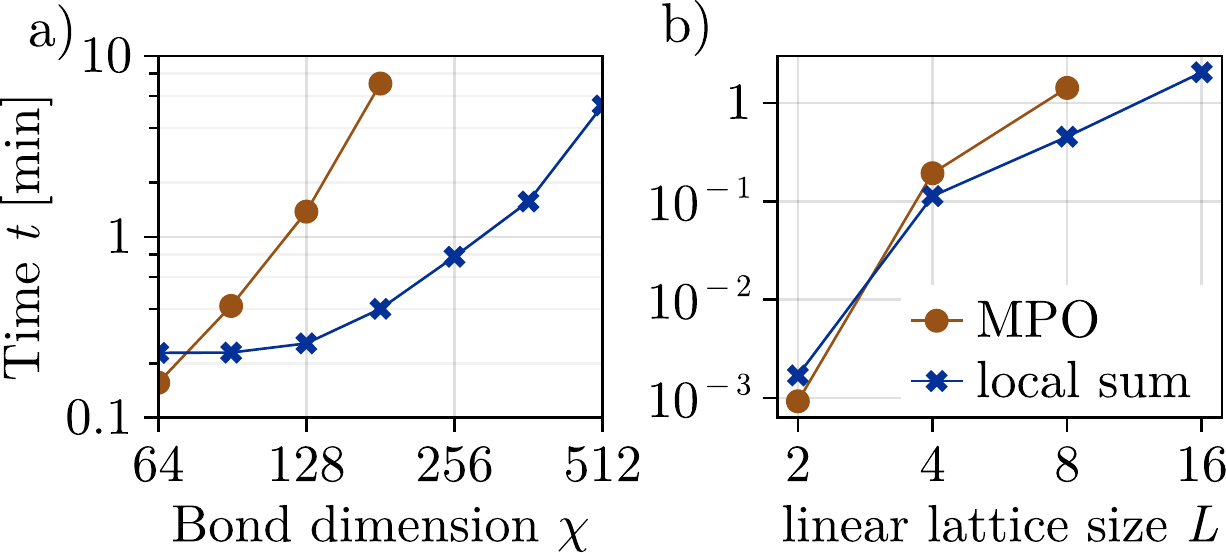}
    \caption{Comparison between the MPO and local sum formalism, showcasing the associated time for a single TDVP step as a function of a) the bond dimension at system size $8\times8$ and b) the linear lattice size at fixed bond dimension $\chi=128$. The comparison was performed on a GPU. Note the missing points for the MPO simulation at the largest bond dimension/system size, which is due to the larger memory footprint of the effective environments, which do not fit on the GPU memory anymore.} \label{fig:mpo_timings}
\end{figure}

A different approach can be found by splitting the Hamiltonian into each local component and summing over the action of these local contributions. 
Similarly, the effective Hamiltonians in \eqref{eq:TDVP_eq3} can be represented as the sum over the effective operators obtained by dressing each individual local term in $\mathcal{H}$ with the isometries. 
The special structure of the tree allows for a reduction of the number of summands appearing in the action by collapsing all $n$-body terms acting solely inside one branch of the tree, see~\cite{Felser_2021} for more details. 
Ultimately this leads to a better scaling of the algorithm, both as a function of the bond dimension, see Fig.~\ref{fig:mpo_timings}a, as well as a function of the linear system size, see Fig.~\ref{fig:mpo_timings}b, providing a speedup of up to 20 compared to the MPO-based approach.
We also want to note that using local sums as opposed to the MPO formalism reduces the memory footprint of storing all environments associated to the effective Hamiltonian, the size of which behave as $\mathcal{O}(N \cdot \chi^2 \cdot K^2)$, with $N$ the number of tensors in the tree and $K$ the bond dimension of the MPO. 
This allows us to access larger bond dimensions and system sizes -- note the missing points of the MPO simulations in Fig.~\ref{fig:mpo_timings} -- without having to move objects between the GPU and CPU. 
\section{Results} \label{sec:results}
In this section we use the presented method to investigate the transverse field Ising model on a two dimensional square lattice, which not only provides a rich test ground for theoretical and numerical analysis of e.g. quantum phase transitions, dynamics etc, but is also readily realizable in modern experiments like Rydberg atom arrays~\cite{Manovitz2024}. 
We start by introducing the model and summarizing some analytical predictions with regards to the perturbative limit in which the interaction strength dominates the system.

This limit is of especial interest, since it allows for reliable benchmarking of said experiments while still exhibiting rich and interesting physics. 
We check the analytical predictions as well as their validity for various initial states.
Finally, we investigate different quench settings, showcasing limitations of the method.

\subsection{Transverse Field Ising Model} \label{sec:tfim}

\begin{figure}[t!]
    \centering
    \includegraphics[width=0.5\linewidth]{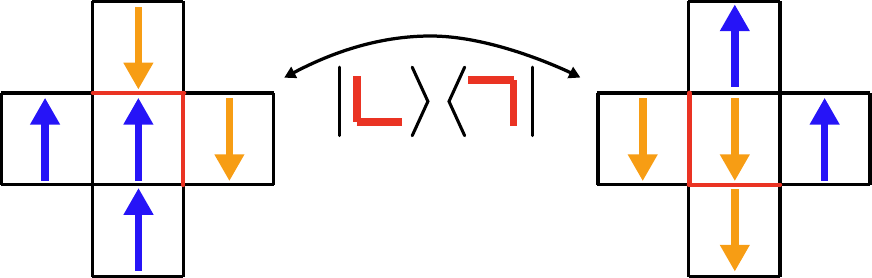}
    \caption{Graphical representation for one of the transitions present in the PXP Hamiltonian~\eqref{eq:PXP}. Only those configurations are allowed, where flipping the center spin does not change the number of domain walls present.} \label{fig:PXP}
\end{figure}

The transverse field Ising model (TFIM) is a paradigmatic model exhibiting a quantum phase transition at $g_c/J \approx 3.04$ and a thermal transition at a temperature of $T_c/J \approx 2.27$ at $g = 0$~\cite{Bloete_2002}.
We consider the model with the addition of a longitudinal field, thus breaking the spin flip symmetry: 
\begin{equation}
\mathrm{H} = - J \cdot \sum_{\langle i,j \rangle} \sigma^x_i \sigma^x_j - \sum_i\Big( g \cdot \sigma^z_i + h \cdot \sigma^x_i\Big), \label{eq:TFI}
\end{equation}
with the first sum running over neighboring spin pairs.
In the infinite interaction coupling limit, i.e. $J \rightarrow \infty$, the emergence of dynamical constraints and Hilbert space fragmentation restrict the allowed transitions to those, which do not change the expectation value of the domain wall length operator
\begin{equation}
\mathrm{D} = \frac{1}{2} \sum_{\langle i,j \rangle} (1 - \sigma^x_i \sigma^x_j). \label{eq:domain_wall_length}
\end{equation}
This concept can be generalized by means of a Schrieffer-Wolff transformation~\cite{Balducci_2022,balducci_interface_2022}, which perturbatively decouples sectors consisting of states with different domain wall lengths, which in zeroth order yields the PXP-Hamiltonian:
\begin{align}
    \mathrm{H}_{\mathrm{PXP}} =
    &- g \Big( \sket{\ulmove}_i{}_i\sbra{\drmove} + \sket{\dlmove}_i{}_i\sbra{\urmove} + \sket{\hhmove}_i{}_i\sbra{\vvmove} + \mathrm{h.c.} \Big).
    \nonumber\\
     &- h \sum_i \sigma_i^z\label{eq:PXP}
\end{align}
The notation can be understood as each bra/ket representing a cross of 5 spins, with the red line denoting a domain wall, i.e. spins separated by a red line are oriented oppositely. Fig.~\ref{fig:PXP} shows a graphical representation for one of the transitions present in the PXP Hamiltonian~\eqref{eq:PXP}.

Starting from the PXP Hamiltonian, analytical predictions~\cite{Balducci_2022,balducci_interface_2022} regarding the dynamics of certain initial conditions can be made, which we investigate in the following. 
\subsection{Strips}
\begin{figure}[t!]
    \centering
    \includegraphics[width=1\linewidth]{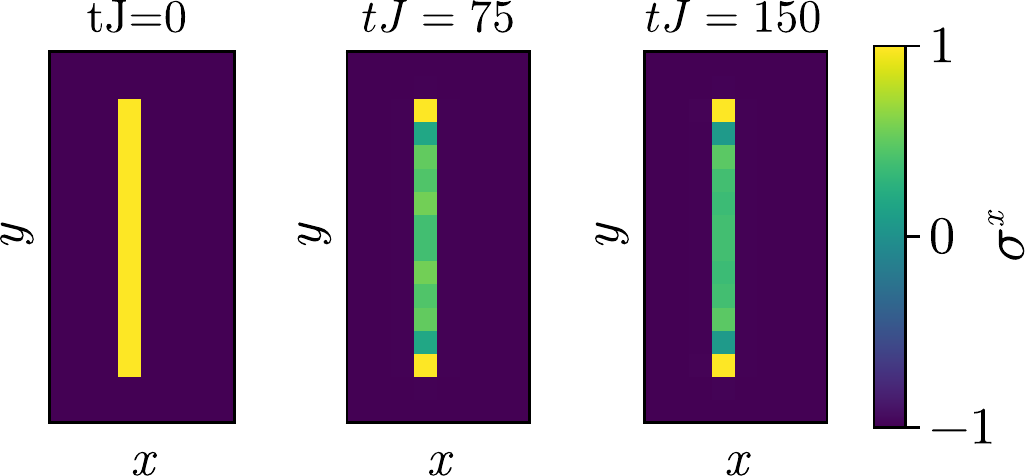}
    \caption{Snapshots of the evolution of a strip of length $l=12$ of up-spins, embedded in a lattice of size $16 \times 8$ of down-spins with $g/J=0.1$, taken at three different times.} \label{fig:stripSnapshots}
\end{figure}

\begin{figure}[t!]
    \centering
    \includegraphics[width=1\linewidth]{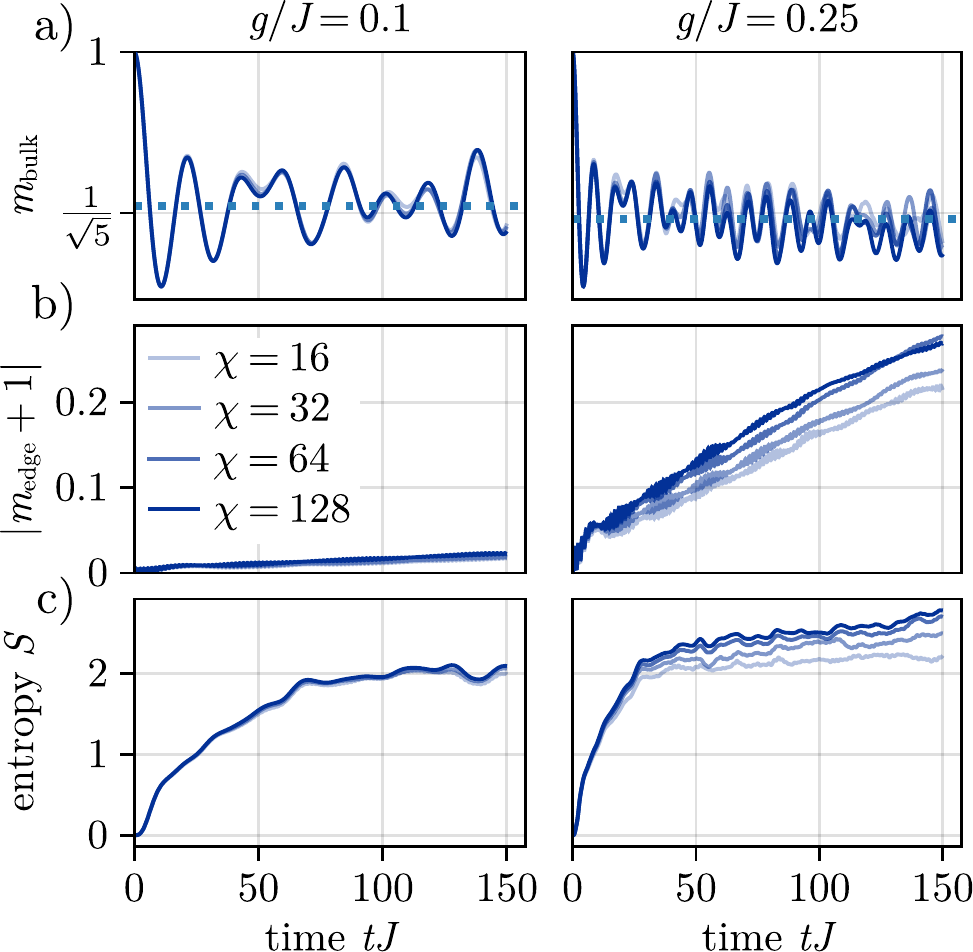}
    \caption{Time evolution of a strip of length $l=12$ of up-spins, embedded in a lattice of size $16 \times 8$ of down-spins for different transverse fields and bond dimensions $\chi =16,32,64,128$. A more translucent line corresponds to a smaller bond dimension. a) Bulk magnetization of the strip, taken as the mean of the 4 innermost spins. It agrees well with the analytical prediction for both magnetic fields. b) The edge magnetization should remain fixed at $1$ in the PXP approximation, which is the case for the smaller transverse field. For the larger one the approximation seems to break down, as the magnetization clearly departs from the expected value. c) Half-system entanglement entropy, cutting the strip in half. For the smaller transverse field, the entanglement entropy is already saturated at a bond dimension $\chi=16$, while it keeps growing for the larger one with increasing bond dimension.} \label{fig:stripTimeEv}
\end{figure}

Starting with a strip of down spins in a sea of up spins, see Figure~\ref{fig:stripSnapshots} for snapshots of the initial and some of the late time states, the PXP-Hamiltonian~\eqref{eq:PXP} only allows transitions of spins that have two up- and two down-neighbors. From that follows that the outer edge spins are fixed to +1, while  inside the strip some transitions are still allowed. In the long time limit, as well as in the limit of first setting the lattice size $L \rightarrow \infty$ and then the strip length $l \rightarrow \infty$, the magnetization of the spins inside the strip approaches~\cite{balducci_interface_2022} 
\begin{equation}
\langle m_\infty \rangle = -1 - \frac{2}{(2 \phi -1) \phi} = \frac{1}{\sqrt{5}}. \label{eq:stripAnalyticalRes}
\end{equation}
with $\phi = (1 + \sqrt{5})/2$ the golden ratio.
Due to the dynamics being strongly restricted to a low-dimensional subspace of the Hilbert space, this can be efficiently simulated using tensor network methods with a low bond dimension. Fig. \ref{fig:stripTimeEv} shows the bulk and edge magnetization as well as the entanglement entropy of a strip of length $l=12$, embedded in a lattice of size $16\times8$. Lines of different opacity denote results obtained for different bond dimensions $\chi =16,32,64,128$, where a more translucent line corresponds to a smaller bond dimension.
For both transverse fields the time average of the bulk magnetization, shown as the dotted line agrees well with the analytical result in Eq.~\eqref{eq:stripAnalyticalRes}, which has been marked on the y-axis. 
However the magnetization of the edge spin clearly departs from the expected value of $m_\mathrm{edge}=-1$ for the larger transverse field $g/J=0.25$, which becomes even more apparent as the bond dimension increases. 
For the smaller transverse field $g/J=0.1$ it deviates by at most $2\%$, even up to simulation times of $tJ=150$.
Such a behavior signifies the breakdown of the PXP approximation somewhere between $0.1<g/J<0.25$. 

This is confirmed by the half-system entanglement entropy, i.e. when cutting the strip in half, shown in the bottom plots. 
For the smaller transverse field, the entanglement entropy stays constant at later times and is already saturated at a bond dimension of $\chi=32$. This behavior is expected, since the dynamics within the highly restrictive PXP-approximation should approach an equilibrium. 
For the larger transverse field, the entanglement entropy grows more rapidly and keeps growing with time and with increasing bond dimension, indicating a non-trivial and less restricted dynamics.

Finally we want to note that the analytical predictions, which were made in the thermodynamic limit for infinitely long strips and in the limit $J\to\infty$, provide an accurate description even at finite system sizes and coupling strengths investigated here. 

\subsection{Corners}

\begin{figure}[t!]
    \centering
    \includegraphics[width=0.9\linewidth]{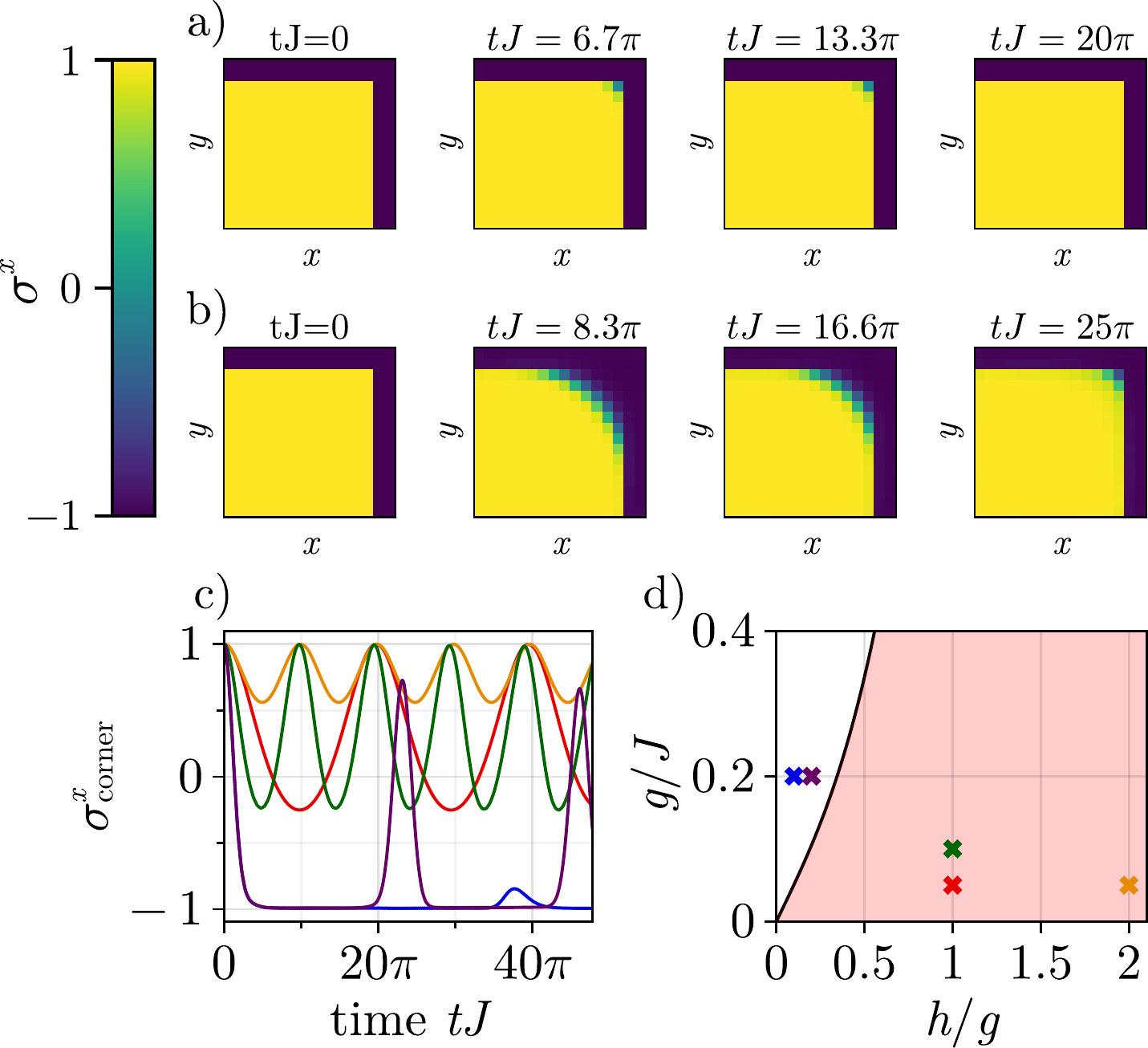}
    \caption{a) and b) Snapshots of the time evolution of a corner state for $g/J=h/J=0.05$ and $g/J=0.2$ and $h/J=0.02$ respectively. c) Time evolution of the magnetization of the corner spin for different combinations of transverse and longitudinal fields with OBC. The color coding of the lines correspond to the values of denoted by crosses of the same color in d). For parameters within the red, localized regime we observe the clean oscillation over several periods with a full recovery of the initial magnetization of $\sigma^x_\mathrm{corner}=1$. The other parameters, i.e. within the white, ergodic regime, show a clear deterioration of the oscillations, indicating the breakdown of the PXP-approximation.} \label{fig:cornerPeriod}
\end{figure}

In the following section we investigate the dynamics of so-called corner states. 
These states have a single interface which can be parametrized as a discrete, Lipschitz-continuous function on the two dimensional lattice, rotated by $45^\circ$. 
For this class of functions, the PXP Hamiltonian can be mapped to a one-dimensional, free fermionic model
\begin{equation}
    H_F= -g \sum_{x \in \mathbb{Z}} (\psi_x^\dag \psi_{x+1} + \mathrm{h.c.}) + 2h \sum_{x \in \mathbb{Z}} x \psi_x^\dag \psi_x, \label{eq:fermHam}
\end{equation}
see~\cite{Balducci_2022,balducci_interface_2022} for more details. 

The mapping is such that within the rotated frame, every up slope of the interface can be associated with a fermion, while every down slope is a vacant site, restricting it to a one-dimensional lattice. 
The only transitions allowed by the PXP Hamiltonian are those that change the orientation of two neighboring, oppositely-oriented slopes, thus effectively moving the associated fermion by one site. 

The presence of a linear potential for the fermionic modes due to the longitudinal field is well known to lead to Wannier-Stark localization~\cite{Emin_1987,balducci_interface_2022}: for $h\neq0$ the dynamics of the fermions is restricted to oscillations around their initial position with a frequency $\nu = \pi \cdot \frac{J}{h}$ and an amplitude $A \propto |\frac{g}{h}|$.

Snapshots of a numerical simulation of that effect are presented in Fig.~\ref{fig:cornerPeriod}a for a corner of size $14 \times 14$ embedded in a lattice of size $16 \times 16$ with open boundary conditions and parameters $g/J=h/J=0.05$. We observe a small oscillation amplitude and a full recovery of the initial state at $tJ = 20\pi$.

For a different set of parameters, $g/J=0.2$ and $h/J=0.02$, shown in in Fig.~\ref{fig:cornerPeriod}b, the initial state is not recovered after the predicted oscillation period $T=25\pi$, indicating a breakdown of the PXP-approximation, to be discussed in the following.

At a finite coupling strength $J \neq \infty$ it is a priori unclear wether Wannier-Stark localization persists. 
By analyzing the level spacings of corner systems to first order in the Schrieffer Wolff perturbation theory, the authors of ~\cite{Balducci_2022,balducci_interface_2022} found a relation between the ergodicity of these systems and parameters of the Hamiltonian. In Fig.~\ref{fig:cornerPeriod}d we show the thereby identified ergodic/non-ergodic regions in red/white, respectively. 
Naively, increasing the transverse field leads to delocalization, while increasing the longitudinal field enhances the Wannier-Stark effect and thus localizes the system.

In Fig.~\ref{fig:cornerPeriod}c we show results for parameter combinations belonging to the different regimes, the color coding corresponding to the crosses in d). 
Although some oscillation can be observed in all systems, there is a clear distinction regarding the quality of the revivals; for systems within the delocalized regime, the corner magnetization shows large deviations from the expected value of $\langle \sigma^x_\mathrm{corner} \rangle=1$ already after one oscillation period.
Furthermore, even for systems from the localized regime we observe a slight deviation of the numerical oscillation frequency from the expected ones, denoted by the tick lines at multiples of $10 \pi$. 
This can be due to a number of factors: on one hand, we are working on a finite lattice, while the predictions were calculated in the thermodynamic limit. 
On the other hand, the PXP approximation is strictly speaking only valid in the limit $g,h \ll J$, while additional terms are present for finite couplings and fields. 

\subsection{Bubbles}
We now turn to initial conditions of finite bubble states of up-spins embedded in a background of down-spins. 
The initial state is an $8\times8$ square of up spins embedded in a background of down spins on a $16 \times 16$ lattice. 

\begin{figure}[hbt!]
    \centering
    \includegraphics[width=0.9\linewidth]{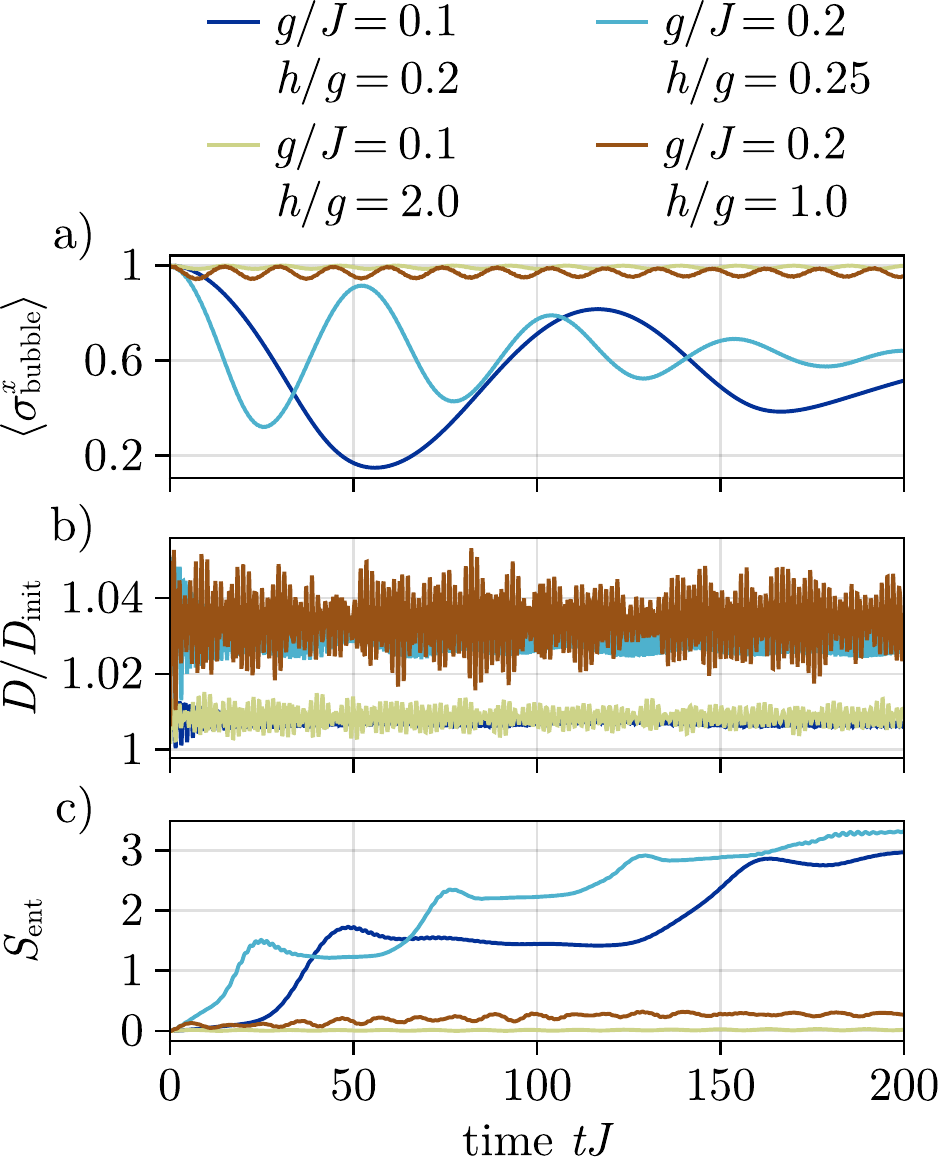}
    \caption{Dynamics for different combinations of transverse and longitudinal fields of initial bubble states of size $8\times8$, embedded in a $16\times16$ lattices. a) shows the mean polarization of the bubble as a function of time. The amplitude and period of the observed oscillations get larger when increasing the ratio of $|g/h|$, which ultimately leads to an observable decay of the oscillation for the light/dark blue lines. b) The domain wall length stays close to its initial value for all parameters, indicating that the dynamics is still restricted to a single domain wall length sector. c) Half system entanglement entropy as a function of time. The regularly oscillating bubbles show little entanglement growth, in contrast to both blue lines, corresponding to decaying bubbles.} \label{fig:bubbles}
\end{figure}

For large enough bubbles and longitudinal fields, the dynamics of the bubble states essentially reduces to that of four individual, oscillating corners, provided the oscillation amplitude, determined by the transverse and longitudinal fields is smaller than half of the linear bubble size. 
Such a description will eventually break down -- using the fermionic description of the corner states, it was predicted~\cite{Balducci_2022,balducci_interface_2022} that the timescale of that process is inversely proportional to the squared probability of finding a fermion at half bubble size, which entails an exponential dependence on the bubble size.

Fig.~\ref{fig:bubbles}a explores this behavior qualitatively on the example of the bubble magnetization for different combinations of the Hamiltonian parameters. We clearly observe different regimes: for larger longitudinal fields $h/g \approx 1$ the dynamics is strongly constrained, resulting in persistent oscillations, which do not deteriorate on the time scales investigated here. The entanglement entropy, shown in Fig.~\ref{fig:bubbles}c stays almost constant throughout the evolution, confirming the restricted nature of the dynamics.

\begin{figure*}[ht!]
    \centering
    \includegraphics[width=\linewidth]{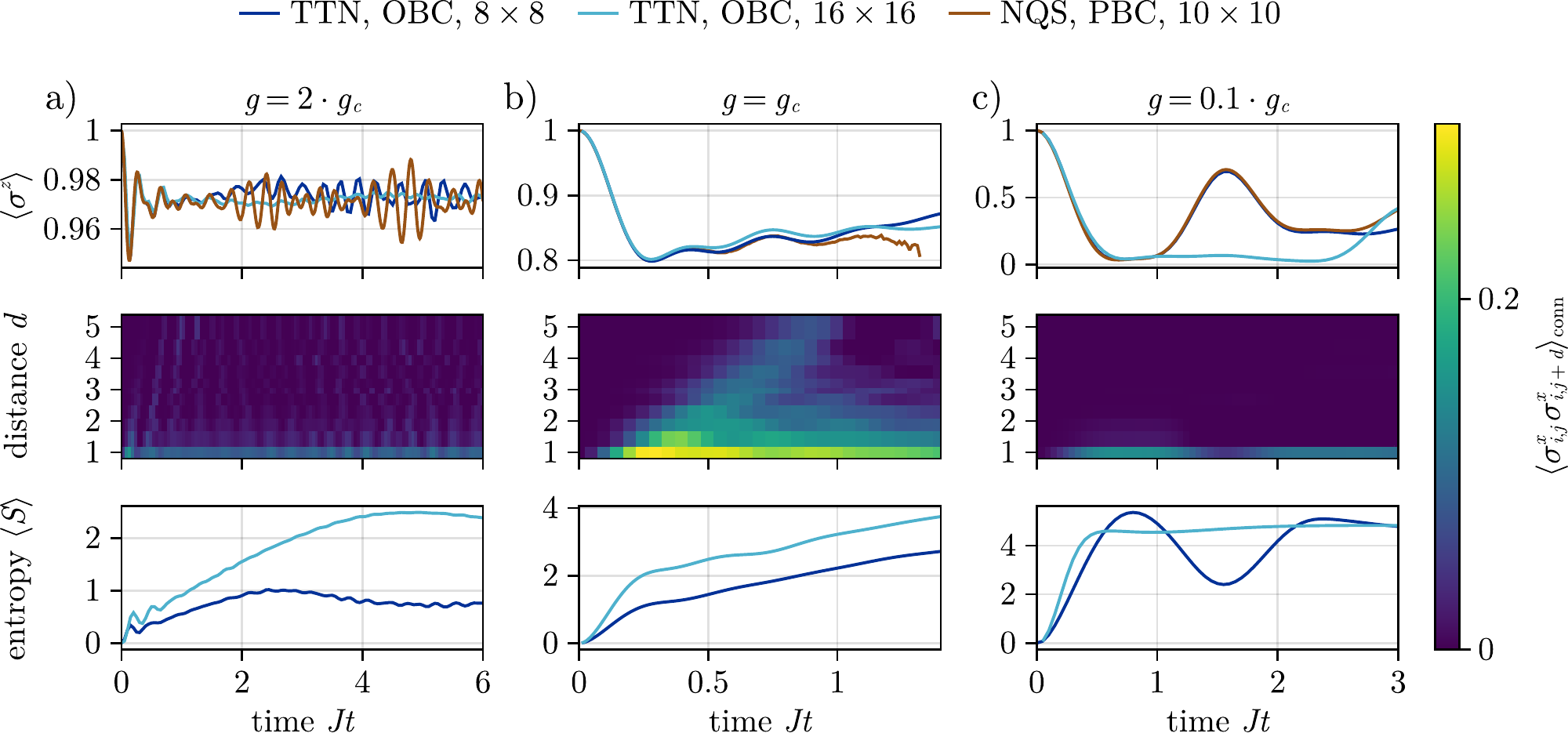}
    \caption{Investigation of three different quenches, starting from a transversely polarized initial state to a) the paramagnetic regime ($g=2 \cdot g_c$), b) the critical point ($g=g_c$) and c) the ferromagnetic regime ($g=0.1 \cdot g_c$). The TTN simulations were performed on a $8 \times 8$ lattice with a bond dimension of $\chi=512$ and $16 \times 16$ lattice with $\chi=362$. On both lattices we imposed OBC. The reference results were obtained with neural quantum states~\cite{Schmitt_2020} (NQS) and PBC on a 10 by 10 lattice. The first row shows the mean magnetization. Results obtained with TTN and NQS disagree only for the $16\times 16$ simulations for the first two quenches due to the large growth of entanglement, see third row. Finite size effects explain minor differences between the simulations, due to correlations spreading in a lightcone and reaching the boundary after a characteristic time, see second row.} \label{fig:quenchBenchmark}
\end{figure*}
On the other hand, for $h/g \approx 0.1$, the observed oscillations are clearly damped, which can be attributed to the amplitude of the corner oscillations being large enough to feel each others presence. 
This destroys the perfect coherence of the oscillations, which is reflected in the rapid growth of entanglement.
Most notably, the growth of the entanglement entropy shows a step-wise behavior, with the position of the steps coinciding with the minimum of the respective oscillation, see~Fig.~\ref{fig:bubbles}a. 
This observation suggests that most of the entanglement growth in the evolution happens in the regime where neighboring corners start to interact with each other.
We want to note that the mean magnetization value the damped oscillations are approaching is not the one corresponding to the fully thermalized state -- the dynamics is still restricted in the Hilbert space sector corresponding to states with the same domain wall length as the initial state. 
Thus, the observed equilibration occurs for the most part within said sector. This is confirmed by Fig.~\ref{fig:bubbles}b, where we see the expectation value of the domain wall length operator remaining close to its initial value during time evolution.

\subsection{Quench dynamics} \label{sec:quench}
As a last application, we investigate the quench dynamics of a transversely polarized initial state, i.e., starting deep in the paramagnetic phase, to $g=2\cdot g_c$, see~Fig.~\ref{fig:quenchBenchmark}a, to the critical point $g=g_c$, see Fig.~\ref{fig:quenchBenchmark}b, and into the ferromagnetic regime $g=0.1\cdot g_c$, see Fig.~\ref{fig:quenchBenchmark}c)
For the benchmark, the results are compared to ones obtained with \textbf{N}eural \textbf{Q}uantum \textbf{s}tates (NQS) and PBC~\cite{Schmitt_2020}, see top row of Fig.~\ref{fig:quenchBenchmark}. 
The TTN simulations were performed on lattices of size $8 \times 8$ with a bond dimension of $\chi=512$  and $16 \times 16$ with $\chi=362$. Note that we imposed OBC, which is more favorable for tensor network simulations as that leads to an overall slower growth of entanglement. 
To enable a quantitative comparison between the two methods and different boundary conditions, we only consider the mean magnetization of the four center spins as the magnetization for our TTN simulation, which reduces unwanted boundary effects.

Starting from Fig.~\ref{fig:quenchBenchmark}a, i.e the quench to $g=2g_c$, the magnetizations coincide between all architectures and lattice sizes, up to timescales in which the finite size of the systems becomes relevant to the dynamics.
These slight deviations can be understood by looking at the second row, which shows how correlations spread through the system.
As soon as they reach the boundary, the choice of the boundary condition affects the dynamics, leading to different solutions for the magnetization. 
The third row shows the behavior of the half-system entanglement entropy with time. 
For the $16\times16$ system the entanglement entropy is roughly doubled compared to the $8\times8$ lattice, which can be attributed to the doubled length of the boundary corresponding to the considered half-system bi-partition. 
For this quench the entanglement entropy does not show any saturation, indicating the reliability of our results.

\begin{figure}[t!]
    \centering
    \includegraphics[width=\linewidth]{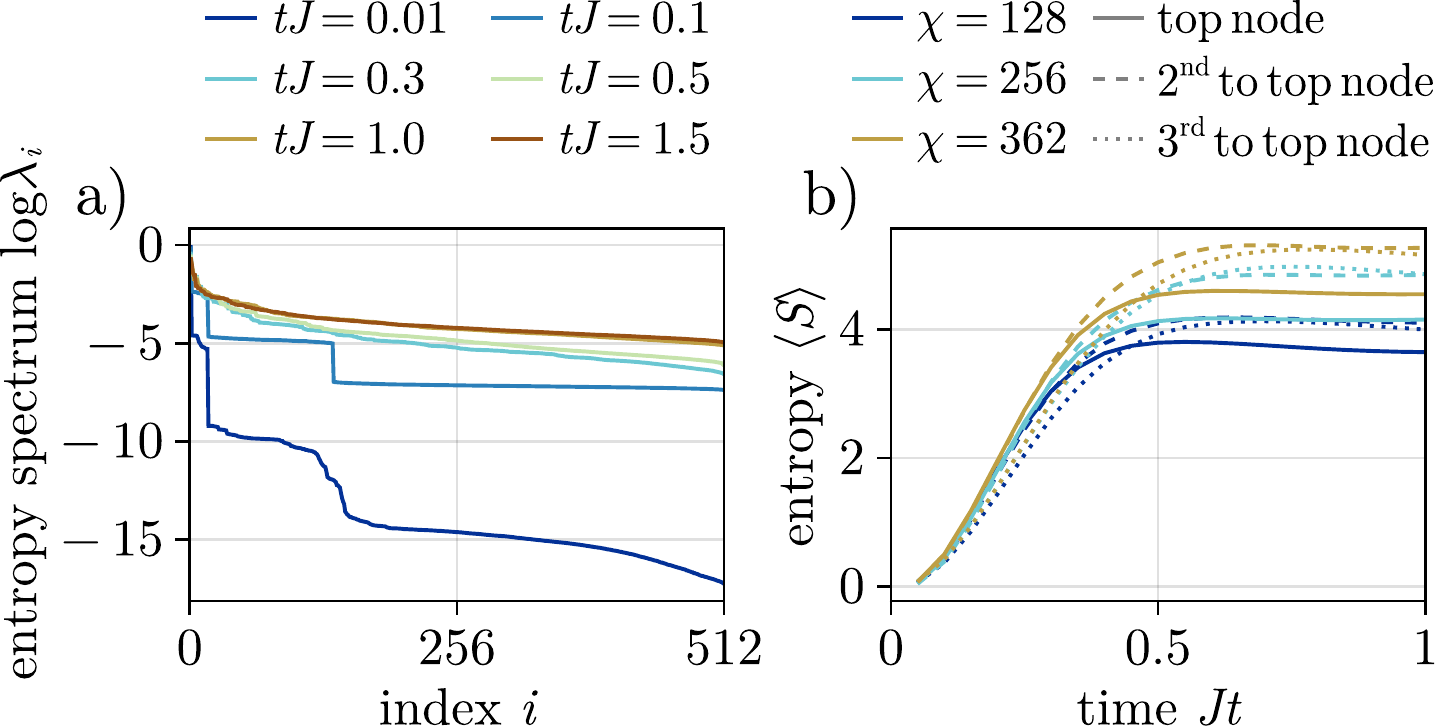}
    \caption{a) Snapshots of the half-system entanglement spectrum for the quench to $g=g_c$ and $8 \times8$ system size. For times $tJ\geq0.3$ the spectrum becomes flat, indicating that important singular values have been cut off due to the finite bond dimension. b) Time evolution of the entanglement entropy for different bond dimensions and bi-partitions on the $16 \times 16$ lattice and the quench to $g=0.1 \cdot g_c$. Cutting along the top/$2^\mathrm{nd}$ to top/$3^\mathrm{rd}$ to top node corresponds to a half-system/$\frac{1}{4}$ to $\frac{3}{4}$/$\frac{1}{8}$ to $\frac{7}{8}$ partition of the lattice, respectively. We observe a saturation of the entanglement entropy already at the $3^\mathrm{rd}$ to top node, signaling a breakdown of the finite bond dimension approximation at these levels.} \label{fig:quenchBenchmark_entropy}
\end{figure}

In Fig.~\ref{fig:quenchBenchmark}b, corresponding to the quench to the critical point, we observe a slight difference in the magnetization between the two system sizes accessed by our TTN simulations -- while results for the $8\times8$ system agree with the NQS results up to time $tJ\approx1$, at which point again correlations (see second row) reach the boundary of the system, results for the $16\times16$ system show slight deviations already at an earlier time. 
In order to better understand this behavior, we investigated the spectrum of the entanglement entropy of the $8\times8$ system in Fig.~\ref{fig:quenchBenchmark_entropy}~a, with a bond dimension of $\chi=512$. While at earlier times the entanglement spectrum clearly exhibits a clear hierarchy of scales, starting from $tJ\approx0.3$ the spectrum becomes more and more flat, resulting in the need for a higher bond dimension to reliably represent the state and a slow convergence with increasing $\chi$. 
This effect can be expected to be even more detrimental for larger systems due to the aforementioned relation between entanglement entropy and the size of the boundary, leading to the observed deviations. 

Finally, for Fig.~\ref{fig:quenchBenchmark}c, i.e. the quench across the critical point into the ferromagnetic regime, the limitations become even more obvious at the larger system size, as already at $tJ\approx1$ the magnetization starts to deviate strongly from the reference. 
This behavior is once again reflected in the entanglement entropy, see Fig.~\ref{fig:quenchBenchmark}c: while the method is still able to reliably capture the dynamics of the magnetization for the $8\times8$ system, the entanglement entropy for the $16\times16$ system is already saturated at $tJ\approx 0.4$. 
By looking at the entanglement entropy at different bond dimensions and levels of the tree, corresponding to different bipartitions of the lattice, we observe that the entanglement entropy already shows saturation at the 3rd level from the top, see Fig.~\ref{fig:quenchBenchmark_entropy}b. 
This corresponds to $1/8$ to $7/8$ bipartition of the lattice. A saturated entanglement at a lower level suggests an even worse behavior at higher levels, 
rapidly saturating the bond dimension and restricting the representation power of the net.
\section{Discussion} \label{sec:discussion}
In this work, we explored the capabilities of TTNs to investigate the non-equilibrium dynamics of a two-dimensional quantum magnet under various aspects.
We provided a comprehensive study of the gain in efficiency by using \emph{GPU} accelerators, as well as local sums over the traditional MPO representation of Hamiltonians, establishing TTNs as a viable and performant member of the tensor network toolboox.
This allowed us to investigate the performance of Tree Tensor Networks applied to solving the time evolution of the two-dimensional quantum Ising model, in both a constrained, perturbative regime as well as for more general quenches, by reaching large system sizes of up to $16 \times 16$ lattice sites and very long time scales. 
The dynamics within the constrained regime can be captured reliably with this method due to the slow growth of entanglement for these systems, such that we are able to not only verify analytical predictions made in the infinite interaction limit, but also expand and investigate effects coming from weakening some of the assumptions, e.g. introducing a finite interaction strength as well as finite lattices.

Going forward, this method can be used to get a more thorough understanding of e.g. the effect of finite couplings on the lifetime of bubbles, which, to our knowledge, has not been fully resolved yet. Such an analysis would furthermore lay the groundwork for studying phenomena such as false vacuum decay. 

Finally, we have investigated three paradigmatic quenches, which have been benchmarked extensively with various methods before. 
While these quenches are generally more challenging due to a faster growth of entanglement, especially close to the phase transition line, it can still be used as a controlled benchmark method for specific cases, e.g. for system sizes up to 8$\times$8 sites, giving access to the entanglement structure and allowing the evaluation of local observables.
This can be seen as a valuable resource considering the rapidly evolving field of quantum simulators, specifically Rydberg atom arrays, for which these simple quenches can provide non-trivial benchmark scenarios.

\begin{acknowledgments}
We acknowledge fruitful discussions with A. Gambassi and thank L. Pavešić and S. Montangero for their comments on the manuscript.
The tree tensor network simulations presented in this work were produced with \rm{TTN.jl}~\cite{Tausendpfund2024}, a software package we developed based on the \rm{ITensor} library~\cite{Fishmann2022}. 
MS and WK were supported through the Helmholtz Initiative and Networking Fund, Grant No. VH-NG-1711.
We acknowledge support from the Deutsche Forschungsgemeinschaft (DFG) under 
Germany's Excellence Strategy - Cluster of Excellence Matter and Light for Quantum Computing (ML4Q) EXC 2004\slash 1 – 390534769 (NT, MR and MS),
under Grant No. 277101999 -- CRC network TR 183 (NT and MR),
under project 499180199 -- FOR 5522 (MH).
MH has received funding from the European Research Council (ERC) under the European Union’s Horizon 2020 research and innovation programme (grant agreement No. 853443).
MR and NT acknowledge funding under Horizon Europe programme HORIZON-CL4-2022-QUANTUM-02-SGA via the project 101113690 (PASQuanS2.1).
The authors gratefully acknowledge the Gauss Centre for Supercomputing e.V. (www.gauss-centre.eu) for funding this project by providing computing time through the John von Neumann Institute for Computing (NIC) on the GCS Supercomputer JUWELS~\cite{JUWELS} and through FZJ on JURECA~\cite{JURECA2021} at J\"ulich Supercomputing Centre (JSC).
Data and code will be made available soon.
\end{acknowledgments}

\appendix

\bibliography{bibliography}

\end{document}